\documentclass[10pt,a4paper,openright,tablecaptionabove]{scrartcl}
\usepackage{subfig}
\usepackage[english]{babel}
\usepackage{amsmath, amsthm, amssymb}
\usepackage{bbm}
\usepackage{url}
\usepackage{tikz-cd}
\usepackage{braket}
\usepackage{path}
\usepackage{nicefrac}
\usepackage{amssymb}
\usepackage{amsmath}
\usepackage{amsthm}
\usepackage{cancel}
\usepackage{indentfirst}
\usepackage{eucal}
\usepackage{appendix}
\usepackage[utf8]{inputenc}
\usepackage{geometry}
\usepackage{verbatim}
\usepackage{nameref}
\usepackage{alltt}
\usepackage{hyperref}
\usepackage{pgfplots}
\usepackage{dsfont}
\usepackage{appendix}
\usepackage[font=small,labelfont=bf]{caption} 

\newtheorem{rem}{Remark}

\newcommand{\hbx}{\hat{\boldsymbol{x}}}
\newcommand{\bx}{\boldsymbol{{x}}}
\newcommand{\hx}{\hat{{x}}}

\newcommand{\hbp}{\hat{\boldsymbol{p}}}
\newcommand{\bp}{\boldsymbol{{p}}}
\newcommand{\hp}{\hat{{p}}}

\areaset[current]{500pt}{680pt}
\begin{document}

\thispagestyle{empty}
\begin{center}
\Large{{\bf Embedding of the Racah Algebra R($\boldsymbol{n}$) and Superintegrability}}
\end{center}
\vskip 0.5cm
\begin{center}
\textsc{$^1$Danilo Latini, $^2$Ian Marquette and $^2$Yao-Zhong Zhang}
\end{center}
\begin{center}
$^1$Via Appia Antica 2, 00045 Genzano di Roma, Rome, Italy\\
$^2$School of Mathematics and Physics, The University of Queensland, Brisbane, QLD 4072, Australia
\end{center}
\vskip 0.5cm

\vskip  1cm
\hrule
\begin{center}
\textsf{{\bf abstract}}
\end{center}
  \begin{abstract}
 	\noindent  The rank-$1$ Racah algebra $R(3)$ plays a pivotal role in the theory of superintegrable systems.  It appears as the symmetry algebra of the $3$-parameter system on the $2$-sphere from which all second-order conformally flat superintegrable models in $2$D can be obtained by means of suitable limits and contractions. A higher rank generalization of $R(3)$, the so-called rank $n-2$ Racah algebra $R(n)$, has been considered recently and showed to be the symmetry algebra of the general superintegrable model on the $(n-1)$-sphere. In the present work, we show that such an algebraic structure naturally arises as embedded inside a larger quadratic algebra characterizing $n$D superintegrable models with non-central terms. This is shown both in classical and quantum mechanics through suitable (symplectic or differential) realisations of the Racah and additional generators. Among the main results, we present an explicit construction of the complete symmetry algebras for two families of $n$-dimensional maximally superintegrable models, the Smorodinsky-Winternitz system and the generalized Kepler-Coulomb system.  For both families, the underlying symmetry algebras are higher-rank quadratic algebras containing the Racah algebra $R(n)$ as subalgebra. High-order algebraic relations among the generators of the full quadratic algebras are also obtained both in the classical and quantum frameworks. These results should shed new light to the further understanding of the structures of quadratic algebras in the context of superintegrable systems.
 \end{abstract}
\vskip 0.35cm
\hrule

%
%
%
%
%
\section{Introduction}
\label{intro}

Superintegrable systems represent a special subset of finite-dimensional integrable Hamiltonian systems. They are characterised by the existence of a number of constants of motion that exceeds the number of degrees of freedom.  Given an autonomous Hamiltonian system $H=H(\bx, \bp)$, Liouville integrability requires the existence of $n$ well-defined functionally independent  functions on the phase-space $H_i=H_i(\bx, \bp)$  such as $\{H, H_i\}=0$ for $i=1, \dots, n$. The Hamiltonian, which trivially commutes with itself, appears in the list as $H_1:=H$. Moreover, the above constants have to be in involution, i.e. $\{H_i, H_j\}=0$ for $i,j =1, \dots, n$.   If besides the  $n$ integrals of motion there exist other $k$ functionally independent additional ones, with $1 \leq k \leq n-1$, the system is called superintegrable. If $k=1$, then we deal with minimally superintegrable (mS) systems whereas if $k=n-1$ with maximally superintegrable (MS) ones. In the latter case, because of the existence of such a large number of symmetries very special properties arise, such as the closure of bounded orbits and the periodicity of the motion in classical mechanics or the phenomenon of  additional (sometimes referred as accidental) degeneracy of the energy levels in the quantum setting, just to cite a few.  If there exist $n-2$ constants of motion, i.e. just one constant less than that required for maximal superintegrability, then the system is said to be quasi-maximally superintegrable (QMS). For an historical and a state-of-art perspective about superintegrable systems we refer the reader to the papers \cite{MillerPostWinternitz2013R, 2014RCD....19..415B}.  

A well-known example of QMS systems in $n$ dimensions is given by the general central force problem, where $H=H(\bx, \bp)=\bp^2/2 + V(r)$, $r:=|\bx|=\sqrt{\bx^2}$, for which the rotational invariance is sufficient to generate a subset of $2n-3$ functionally independent constants coming from the entire set of $\binom{n}{2}$ $\mathfrak{so}(n)$ generators $L_{ij}=x_i p_j-x_j p_i$. Among the central force potentials there exist just two subcases that turn out to be MS, and they correspond to the following choices: $V_1=V_\textsf{HO}(r) =(1/2) \omega^2 r^2$ and $V_2=V_\textsf{KC}(r) =-\mu/r$.  In three dimensions, this result reflects the validity of the \emph{Bertrand Theorem} \cite{Bertrand1873}, which states that among all the central force potentials in 3D Euclidean space, there exist only two cases for which all bounded orbits are closed, and they correspond to the oscillator and Kepler potentials above, a result that has been also extended to spherically symmetric curved spaces \cite{Ballesteros_2009_}. Their maximal superintegrability in $n$D is ensured thanks to the existence of two additional constants of motion (of the second-order in the momenta) respectively, the (symmetric) Demkov–Fradkin tensor  \cite{Dem, doi:10.1119/1.1971373}, whose $\binom{n+1}{2}$ components $D_{ij}=p_i p_j +\omega^2 x_i x_j$ turn out to be conserved for $V=V_1$ and the Laplace–Runge–Lenz (LRL) vector $\boldsymbol{A}$ \cite{doi:10.1119/1.9745, doi:10.1119/1.10202, goldstein2002classical}, whose $n$ components $A_i=\sum_{j=1}^n (L_{ij}p_j)-\mu x_i/r$ are conserved for $V=V_2$. These additional integrals lead to higher symmetry algebras, i.e. the Lie algebra of the $SU(n)$ group for the isotropic harmonic oscillator and the Lie algebra of the $SO(n+1)$ group (when restricted to the subspace $H=E<0$) for the Kepler-Coulomb system. In general, the integrals of motion of superintegrable systems close to more complicated algebras, which in the $n$D case usually are higher rank polynomial algebras \cite{Hoque_2015, 1751-8121-49-12-125201, Hoque_2015_, Iliev_2017, 10.1088/1751-8121/aac111, Latini_2019}. 

Interestingly enough, when additional non-central terms appear in the Hamiltonian, i.e.  for Hamiltonian of the form $H=H(\bx, \bp)=\bp^2/2+\sum_{j=1}^n a_j/2x_j^2+V(r)$, even if the rotational symmetry is broken due to the presence of the non-central terms, the quasi-maximal superintegrability in $n$D is kept alive thanks to existence of the $\binom{n}{2}$ (second-order) constants of the type $Q_{ij}= L_{ij}^2+a_i x_j^2/x_i^2+a_j x_i^2/x_j^2$, with $Q_{ij}=Q_{ji}$. Analogously to the $\mathfrak{so}(n)$ generators, these new constants cannot be all functionally independent when we raise the dimensions. However, it is possible to extract from them a subset of $2n-3$ functionally independent ones. A good explanation of this can be found in the framework of \emph{coalgebra symmetry approach to superintegrability} \cite{Ballesteros1996, 0305-4470-31-16-009, 1742-6596-175-1-012004}, where a subset of $2n-3$ functionally independent constants automatically arise as the image, under a given symplectic realisation, of the so-called left and right Casimirs of the $\mathfrak{sl}(2,\mathbb{R})$ coalgebra, which in the case of spherically symmetric systems appear as the quadratic Casimirs associated to some rotation subalgebras of $so(n)$, whereas in the presence of non-central terms appear as suitable linear combination of the $Q_{ij}$. The general idea behind such an approach to (super)integrability is to reinterpret 1D dynamical Hamiltonian systems as the images, under a given realisation, of some (smooth) functions of the coalgebra generators. Then, by using the coproduct, the method consists in extending the 1D system to higher dimensions. This is in order to obtain a multi-dimensional version of the original 1D Hamiltonian. The main point is that the higher-dimensional system will result endowed, by construction, with constants of motion arising from the coalgebras left and right Casimirs, which at fixed realisation will result in two pyramidal sets composed by $n-1$ constants of motion (one of them being in common) in involution. The union of the above sets results in a unique set composed by $2n-3$ functionally independent constants.  The construction can be extended also to the quantum case, where Poisson brackets are replaced by commutators and in place of symplectic realisations one considers realisations of the algebra given in terms of differential operators (examples can be found in  \cite{Latini_2019, BALLESTEROS20112053, Riglioni_2013, Riglioni_2014, Post_2015, LATINI20163445}). 

What we are interested to show in this paper is that, in complete analogy to radially symmetric systems where the rotational symmetry $\mathfrak{so}(n)$ is sufficient to achieve quasi-maximal superintegrability in any dimension $n$, when $n$ non-central additional terms appear in the Hamiltonian, quasi-maximal superintegrability is still preserved thanks to the existence of the quadratic constants $Q_{ij}$, and the latter close under commutation to give the well-known generalized Racah algebra $R(n)$. Such an algebraic structure is known in the literature to be the symmetry algebra of the generic superintegrable model on the $(n-1)$-sphere (see \cite{bie2020racah}  and references therein) and pseudo-sphere \cite{Kuru_2020}. It represents a higher-rank generalisation of the rank 1 Racah algebra $R(3)$, which is the symmetry algebra of the generic superintegrable model on the 2-sphere \cite{Kalnins_2007, 2011SIGMA...7..051K, Genest2014, Gaboriaud_2019}. The importance of this model, known as $S9$, relies to the fact that all second-order $2$D superintegrable systems, which have been completely classified  on conformally flat spaces \cite{Kalnins05_,Kalnins05_2,  Kalnins06_}, can be obtained from it by suitable limits and contraction procedures \cite{Kalnins_2013}.

Here, we take a slightly different point of view. In particular, we show that  when a specific form of the potential function $V=V(r)$ is chosen to give MS subcases, the generalized Racah algebra $R(n)$ automatically appear as embedded inside a larger symmetry algebra with additional generators, the latter coming from the additional constants of motion/quantum integrals related to the specific choice of the potential. Two well-known $n$D MS models we discuss in detail in this paper to show this result, i.e. the $n$D Smorodinsky-Winternitz system \cite{PhysRevA.41.5666, FRIS1965354, Makarov1967, EVANS1990483, doi:10.1063/1.529449, Ballesteros_2004}, which is second-order superintegrable, and the $n$D generalized (sometimes called \textquotedblleft extended\textquotedblright) Kepler-Coulomb Hamiltonian \cite{doi:10.1063/1.2840465,1751-8121-42-24-245203, 2011SIGMA...7..054T}, which is instead fourth-order superintegrable. Both the classical and the quantum frameworks will be considered and the obtained results will be constantly compared throughout the paper. 

We organize the work as follows:
\begin{itemize}
	\item In section \ref{sec2}  we investigate the quasi-maximal superintegrability of a family of $n$D Euclidean Hamiltonian systems with non-central terms. This is obtained as a consequence of the existence of an hidden $\mathfrak{sl}(2,\mathbb{R})$ coalgebra symmetry of the model, a result already known in the literature also for more general Hamiltonians defined on $n$D spherically symmetric curved spaces \cite{2009AnPhy.324.1219B}. In this context, we show that the generalized Racah algebra $R(n)$ arises as the quadratic Poisson/associative algebra generated by the constants of motion/quantum integrals of the  $n$D QMS family above. This is achieved through explicit symplectic and differential realisations of the Racah generators, which appear as the building blocks of the left and right Casimirs commonly encountered in the coalgebra symmetry approach to superintegrability. Thus, as a byproduct, we also shed some light on the existing connection among the Racah generators and the left and right Casimirs of the coalgebra in the given realisation. Finally, we explicitly write down high-order structure equations relating the generators of the quadratic algebra both in the classical and quantum case.
	
	\item In section \ref{sec3} we specialise the functional form of the potential $V=V(r)$ in such a way to consider two well-known $n$D MS models with non-central terms, appearing as particular subcases of the QMS family introduced in the previous section. In particular, we analyse the $n$D Smorodinsky-Winternitz and the $n$D generalized Kepler-Coulomb system. For both models, we show that the Racah algebra $R(n)$ appears as embedded inside their symmetry algebra, which is generated by the Racah generators together with the additional ones which arise for the specific choice of the potential function. Furthermore, we present the high-order structure equations also for the full quadratic symmetry algebra, once again both in the classical and the quantum framework.
	
	\item In section \ref{conclusions} we summarize the results of the work by providing some concluding remarks and open perspectives.
\end{itemize}

\section{Quasi-maximal superintegrability and the generalized Racah algebra $\boldsymbol{R(n)}$}
\label{sec2}

Let us consider the classical Hamiltonian function describing a particle of unit mass on the Euclidean $n$-space, under the influence of a central potental $V = V (r)$ and with additional non-central terms breaking the radial symmetry:
\begin{equation}
H= H(\bx, \bp) = \frac{1}{2}\biggl(\bp^2+\sum_{j=1}^n \frac{a_j}{x_j^2}\biggl) \, + \, V\bigl(r\bigl) \, ,
\label{eq:H}
\end{equation}
where $\bx = (x_1, \dots, x_n) \in \mathbb{R}^n$, $\bp = (p_1, \dots, p_n) \in \mathbb{R}^n$  are canonical coordinates and momenta satisfying:
\begin{equation}
\{x_i,x_j\}=\{p_i,p_j\}=0  \, ,\quad  \{x_i,p_j\}=\delta_{ij} \, .
\label{eq:Heis}
\end{equation}
Here $\{f, g\}:= \sum_{i=1}^n (\partial_{x_i}f\partial_{p_i}g-\partial_{p_i}f\partial_{x_i}g)$ are the canonical Poisson brackets in the $n$D configuration space, $V=V(r)$ is any smooth function of its argument $r:=|\bx|=\sqrt{\bx^2}$ and the $n$ parameters $a_j$ are assumed to be real and positive. At a fixed $n \geq 2$, this Hamiltonian is endowed with the following  $\binom{n}{2}$ constants of motion:
\begin{equation}
Q_{ij}= L_{ij}^2+a_i \frac{x_j^2}{x_i^2}+a_j \frac{x_i^2}{x_j^2}  \qquad \quad (1 \leq i<j \leq n) \, ,
\label{eq:Qij}
\end{equation}
where $L_{ij}=x_i p_j-x_j p_i=-L_{ji}$ are the $\binom{n}{2}$ (antisymmetric) generators of $\mathfrak{so}(n)$. This can be directly checked by verifying that $\dot Q_{ij}=\{Q_{ij},H\}=0$. 
Here, and throghout the paper, we use the notation $\partial_{y_i} = \partial/\partial y_i$.
\begin{rem}
	For $n=3$ the Hamiltonian \eqref{eq:H}	appears in the Evans classification \cite{PhysRevA.41.5666}, where the author listed all superintegrable systems in three degrees of freedom possessing constants of motion that are linear or quadratic in the momenta. It can be found in the table of minimally superintegrable systems. We underlying that, in three dimensions, the notions of minimal superintegrability and quasi-maximal superintegrability coincide being $n+1=2n-2$ for $n=3$.
\end{rem}

Notice that, if all the non-central terms disappear, the Hamiltonian becomes  rotationally invariant and the constants of motion \eqref{eq:Qij} collapse to the square of the angular momenta $L^2_{ij}$. In that case, the angular momenta itself are conserved quantities and therefore the Hamiltonian underlies an $\mathfrak{so}(n)$ symmetry. Now, the main point is that the Hamiltonian \eqref{eq:H} can be expressed in terms of the generators of the Poisson-Lie (co)algebra $\mathfrak{A}= \mathfrak{sl}(2,\mathbb{R}) \simeq \mathfrak{su}(1,1)$. In this regard, let us consider the following 1D symplectic realisation for the generators\footnote{Here  $J_{\sigma}^{[1]} := D(J_\sigma)$ ($\sigma=\pm,3$), where $D$ is the symplectic realisation.}:
\begin{equation}
J_+^{[1]} = \frac{1}{2}\bigl(p_1^2+\frac{a_1}{x_1^2} \bigl)\qquad J_-^{[1]} = \frac{1}{2}x_1^2\qquad J_3^{[1]} = \frac{1}{2}x_1 p_1 \, ,
\label{gencla}
\end{equation}
such as:
\begin{equation}
\{J_-^{[1]},J_+^{[1]}\}= 2 J_3^{[1]} \quad \{J_3^{[1]}, J_{\pm}^{[1]}\}=\pm  J^{[1]}_\pm \, ,
\label{eq:classgen}
\end{equation}
where $\{f,g\}:= \partial_{x_1}f\partial_{p_1}g-\partial_{p_1}f\partial_{x_1}g$. The Casimir element is given by: $C=J_3^2-J_+J_-$, in the given realisation:
\begin{equation}
C^{[1]}=(J_3^{[1]})^2-J_+^{[1]}J_-^{[1]}=-a_1/4 \, .
\label{eq:cas1D}
\end{equation}

To rise the dimensionality and reach the $n$ dimensions we can apply the $n$-th coproduct $\Delta^{[n]}: \mathfrak{A} \to \mathfrak{A}  \otimes \dots ^{n)}\otimes  \mathfrak{A}$:
\begin{equation}
\Delta^{[n]}:=\left(\text{id} \otimes \dots^{n-2)} \otimes \text{id} \otimes \Delta^{[2]}\right)\circ \Delta^{[n-1]}  \, ,\qquad \Delta^{[1]}:= \text{id} \, ,
\label{ncopr}
\end{equation}
where the (primitive) coproduct $\Delta^{[2]}: \mathfrak{A} \to \mathfrak{A} \otimes \mathfrak{A}$ acts on the basis generators as follows:
\begin{equation}
\Delta^{[2]}(J_{\sigma})=J_\sigma \otimes 1 + 1 \otimes J_{\sigma} \, , \quad \Delta^{[2]}(1)=1 \otimes 1 \, \quad (\sigma=\pm,3) \, .
\label{eq:primelcop}
\end{equation}
The extension to any monomial in the Universal Enveloping Algebra (UEA) $U(\mathfrak{A})$ is obtained from to the homomorphism property of the coproduct map $\Delta^{[2]}$, i.e. $\Delta^{[2]}(A \cdot B)=\Delta^{[2]}(A) \cdot \Delta^{[2]}(B)$, which extends also to $\Delta^{[n]}$ \cite{0305-4470-31-16-009}. Now, by applying $\Delta^{[n]}$ on the basis generators we get at a fixed realisation\footnote{In the $n$D case we consider a \textquotedblleft $n$-particle\textquotedblright symplectic realisation, i.e.  $J^{[n]}_{\sigma} := (D \otimes D \dots^{n)} \otimes D) (\Delta^{[n]}(J_{\sigma}))$.}:
\begin{equation}
J_+^{[n]} = \frac{1}{2}\bigl(\bp^2+\sum_{j=1}^n\frac{a_j}{x_j^2}\bigl) \qquad J_-^{[n]} = \frac{1}{2}\bx^2 \qquad J_3^{[n]} =\frac{1}{2}\bx \cdot \bp \, ,
\label{eqngen}
\end{equation}
together with:
\begin{equation}
\{J_-^{[n]},J_+^{[n]}\}=2 J_3^{[n]} \quad \{J_3^{[n]}, J_{\pm}^{[n]}\}=\pm  J^{[n]}_\pm \, ,
\label{eq:classgennD}
\end{equation}
where $\{f,g\}:= \sum_{j=1}^n\bigl(\partial_{x_j}f\partial_{p_j}g-\partial_{p_j}f\partial_{x_j}g\bigl)$. At this point, we observe that the Hamiltonian \eqref{eq:H} arises as the following function of the generators:
\begin{equation}
H=J_+^{[n]}  \, + \, V\bigl(\,\bigl(2 J_-^{[n]}\bigl)^{1/2}\,\bigl) = \frac{1}{2}\biggl(\bp^2+\sum_{j=1}^n \frac{a_j}{x_j^2}\biggl)\,+\,V(|\bx|)\, ,
\label{eq:HJ}
\end{equation}
where it should be understood that $H=H^{[n]}$. The total Casimir, in the given realisation, reads:
\begin{align}
C^{[n]}= (J_3^{[n]})^2-J_+^{[n]} J_-^{[n]}&=-\frac{1}{4}\biggl(\sum_{1 \leq i< j}^n \bigl(L_{ij}^2+a_i \frac{x_j^2}{x_i^2}+a_j \frac{x_i^2}{x_j^2}\bigl)+\sum_{i=1}^n a_i\biggl) \nonumber\\
&=-\frac{1}{4}\biggl(\sum_{1 \leq i< j}^n L_{ij}^2+\boldsymbol{x}^2 \sum_{j=1}^n \frac{a_j}{x_j^2}\biggl) \nonumber \\
&=\sum_{1 \leq i< j}^n C_{ij}-(n-2)\sum_{i=1}^nC_{i} \nonumber \\
&=\sum_{1 \leq i< j}^n P_{ij}+\sum_{i=1}^nC_{i}.
\label{eq:total}
\end{align}
where we introduced the classical quantities:
\begin{equation}
C_{ij}:=-\frac{1}{4}\biggl( L_{ij}^2+a_i \frac{x_j^2}{x_i^2}+a_j \frac{x_i^2}{x_j^2} +a_i+a_j\biggl) \, ,\qquad C_{i}:= -\frac{a_i}{4} \, ,\qquad 
P_{ij}:=C_{ij}-C_i-C_j \, .
\label{cijc}
\end{equation}

Thus, we see that the Casimir function, at the $n$-dimensional level, arises as a linear combination of the $P_{ij}$ and $C_i$, the latter being just constants in the given realisation. Clearly, the $\binom{n}{2}$ elements $P_{ij}$ cannot be all functionally independent in any dimension. However, a functionally independent subset immediately arises as a consequence of the fact that the Euclidean Hamiltonian \eqref{eq:H} is endowed with a coalgebra symmetry \cite{Ballesteros1996, 0305-4470-31-16-009, 1742-6596-175-1-012004}. Thus, we can construct the so-called left and right Casimirs of the coalgebra that, at a fixed realisation, read:
\begin{equation}
C^{[m]}=\sum_{1 \leq i< j}^m P_{ij}+\sum_{i=1}^m C_i \qquad C_{[m]}=\sum_{n-m+1 \leq i< j}^n P_{ij}+\sum_{i=n-m+1}^n C_i \qquad   (m=1, \dots, n) \, ,
\label{lr}
\end{equation}
where $C^{[1]}=C_1$, $C_{[1]}=C_n$ are just constants and $C^{[n]}=C_{[n]}$ is the total Casimir \eqref{eq:total}. As a consequence of the underlying coalgebra symmetry the set $U  = U_1\, \cup \,U_2$ with  $U_1 := \{C_{[m]}\}$, $U_2:=\{C^{[m]}\}$ for $m=1, \dots, n$ provides $2n-3$  functionally independent constants of motion. Moreover, since the elements in each subset $U_i$ ($i=1,2$) Poisson commute each other, they define two abelian Poisson subalgebras composed by $n-1$ elements in involution (remember that $C_{[1]}$ and $C^{[1]}$ are just constants and that $C_{[n]}=C^{[n]}$). So, the Hamiltonian \eqref{eq:H} is QMS for any choice of $V=V(r)$. Furthermore, each of the two sets $V_i :=U_i\, \cup \,\{H\}$ ($i=1,2$) is composed by $n$ functionally independent involutive constants of motion, thus leading to multi-integrability of \eqref{eq:H}.
\begin{rem}
	The Euclidean Hamiltonian \eqref{eq:H} appears to be a specific subcase of the more general one discussed in \cite{2009AnPhy.324.1219B}. There, the authors used the hidden $\mathfrak{sl}(2, \mathbb{R})$ coalgebra symmetry of the model to construct the integrals of motion of an Hamiltonian describing a particle of unit mass on a generic $n$D spherically symmetric curved space, under the influence of a central potental $V = V (r)$ and with additional monopole-type and non-central terms, the latter breaking the radial symmetry. As a matter of fact, this result holds true for any Hamiltonian expressed in terms of a smooth function of the generators $H=H(J_+^{[n]}, J_-^{[n]},J_3^{[n]})$, that is endowed with the same $2n-3$ functionally independent constants of motion \eqref{lr},  the left and right Casimirs of the coalgebra  \cite{Latini_2019, 1742-6596-175-1-012004,  2009AnPhy.324.1219B, Ballesteros_2006}.
\end{rem}

\noindent Now, the connection with the Racah algebra $R(n)$ is made explicit by considering the following \textquotedblleft classical  generators\textquotedblright:
\begin{equation}
P_{ij}=-\frac{1}{4}Q_{ij}\, ,
\label{genRacah}
\end{equation}
the \textquotedblleft building blocks\textquotedblright of the left and right Casimirs, together with the three indices ones:
\begin{equation}
F_{ijk}:=\frac{1}{2}\{P_{ij},P_{jk}\} \qquad (i \neq j \neq k) \, .
\label{eq:f}
\end{equation}
With the above definitions, for $i,j,k,l,m,r \in \{1,\dots, n\}$ all different, we have:
\begin{equation}
\{P_{ij},H\}=\{F_{ijk},H\} =0\, , \qquad  \{P_{ij},P_{kl}\}=0 \, , \qquad \{P_{ij}, P_{ik}+P_{jk}\}=0
\end{equation}
and we get a quadratic Poisson algebra whose defining relations are those of the Racah algebra $R(n)$ \cite{bie2020racah}:
\begin{align}
&\{P_{ij},P_{jk}\}=2 F_{ijk} \hskip 5.2cm (F_{ijk}=-{F_{jik}}=-F_{ikj}) \label{eq:PoissonRacah1}\\
&\{P_{jk}, F_{ijk}\}=P_{ik}P_{jk}-P_{jk}P_{ij}+2P_{ik}C_j-2P_{ij}C_k\\
&\{P_{kl}, F_{ijk}\}=P_{ik}P_{jl}-P_{il}P_{jk}\\
&\{F_{ijk}, F_{jkl}\}=F_{jkl}P_{ij}-F_{ikl}(P_{jk}+2C_j)-F_{ijk}P_{jl}\\
&\{F_{ijk}, F_{klm}\}=F_{ilm}P_{jk}-P_{ik}F_{jlm}
\label{eq:PoissonRacah5}
\end{align}
together with $\{F_{ijk}, F_{lmr}\}=0$. This means that the Euclidean Hamiltonian \eqref{eq:H}, for any suitable choice of the potential function $V=V(|\bx|)$, turns out to be endowed with constants of motion which close to give the quadratic algebra $R(n)$. These constants are related to the left and right Casimirs as in \eqref{lr}, so the latter appear in the given realisation as specific linear combinations of them.
\begin{rem}
	This results should be compared with the case of rotationally invariant systems, i.e. when all the non-central terms disappear from the Euclidean Hamiltonian \eqref{eq:H}. As we already mentioned, in that case the components of the angular momentum span the Poisson-Lie algebra $\mathfrak{so}(n)$ and the left and right Casimirs arise as quadratic combinations of the $L_{ij}$, i.e.  $C^{[m]} \sim \sum_{1 \leq i< j}^n L_{ij}^2$ and $C_{[m]} \sim \sum_{n-m+1 \leq i< j}^n L_{ij}^2$.  Thus, the effect of the non-central terms appearing in the Hamiltonian is to break such a rotational symmetry. As a consequence,  the Poisson-Lie algebra $\mathfrak{so}(n)$ is replaced by the quadratic Poisson algebra $R(n)$.
\end{rem}
The closure of the polynomial algebra is obtained by observing that the following high-order relations among the generators hold:
\begin{align}
&F_{ijk}^2-C_i P_{jk}^2-C_j P_{ik}^2-C_k P_{ij}^2+P_{ij}P_{jk}P_{ik}+4 C_i C_j C_k=0 \label{eq:ho1}\\
& 2F_{ijk}F_{jkl}-P_{il}P_{jk}^2+P_{ij}P_{jk}P_{kl}+P_{ik}P_{jk}P_{jl}-2C_j P_{ik}P_{kl}-2C_k P_{ij}P_{jl}+4C_jC_kP_{il}=0\\
& 2F_{ijk}F_{klm}-P_{il}P_{jk}P_{km}-P_{ik}P_{jm}P_{kl}+P_{im}P_{jk}P_{kl}+P_{ik}P_{jl}P_{km}-2C_kP_{im}P_{jl}+2C_kP_{il}P_{jm}=0\\
& 2F_{ijk}F_{lmr}-P_{il}P_{jr}P_{km}-P_{ir}P_{jm}P_{kl}-P_{kr}P_{im}P_{jl}+P_{im}P_{jr}P_{kl}+P_{ir}P_{jl}P_{km}+P_{il}P_{jm}P_{kr}=0 \, .
\label{eq:ho4}
\end{align}

 These cubic algebraic relations can be interpreted as sixth order constraints in terms of explicit realizations involving polynomials of momenta. Let us remark that the notion of algebraic closure at sixth order integrals had been discussed in the classification of two and three dimensional systems on conformally flat spaces \cite{Kalnins05_, doi:10.1063/1.2037567}. Here, we have obtained the high-order relations for the closure of the polynomial algebra in the context of $n$-dimensional superintegrable systems.

Once the connection of the Hamiltonian with the generators of the Poisson-Lie algebra is found, we can construct a quantum analog of it by considering a realisation of $\mathfrak{A}$ given in terms of differential operators \cite{BALLESTEROS20112053, Riglioni_2013, Riglioni_2014, Post_2015, LATINI20163445}. The latter can be constructed by considering the quantum analog of the realisation given in \eqref{gencla}. This can be achieved by replacing the canonical coordinates and momenta with their quantum versions and by considering their standard differential coordinates representation acting on the space of wavefunctions, namely:
\begin{equation}
\hat{x} = x \qquad \hat{p} = - \imath \hbar \partial_x \, ,
\label{eq:qca}
\end{equation}
such as:
\begin{equation}
[\hat{x}, \hat{x}]=[\hat{p}, \hat{p}]=0  \, ,\quad [\hat{x}, \hat{p}]=\imath \hbar \, \mathds{1} \, .
\label{eq:Heisb}
\end{equation}
In this way, we are able to write the new quantum realisation at the 1D level, the quantum analog of \eqref{gencla}, as\footnote{Here we are using a differential realisation $\bar{D}$, i.e. $\hat{J}_{\sigma}^{[1]} := \bar{D}(J_\sigma)$.}:
\begin{equation}
\hat{J}_+^{[1]} =  \frac{1}{2}\bigl(-\hbar^2\partial_{x_1}^2+\frac{a_1}{x_1^2} \bigl)\quad \hat{J}_-^{[1]} = \frac{1}{2}x_1^2  \quad \hat{J}_3^{[1]} =  -\frac{\imath \hbar}{2} (x_1 \partial_{x_1}+1/2) \, .
\label{qreal}
\end{equation}
Here, the classical term $x p$ appears as the symmetrized operator $\frac{1}{2}\{\hat{x},\hat{p}\}$, $\{\hat{A}, \hat{B}\}=\hat{A}\hat{B}+\hat{B}\hat{A}$ being the anti-commutator. The quantum analog of the Poisson relations \eqref{eq:classgen} are now realised in terms of the commutators:
\begin{equation}
[\hat{J}_-^{[1]}, \hat{J}_+^{[1]}]= 2\imath\hbar  \hat{J}_3^{[1]} \quad [\hat{J}_3^{[1]}, \hat{J}_{\pm}^{[1]}]=\pm  \imath \hbar \hat{J}_\pm^{[1]} \, .
\label{eq:quantumgen}
\end{equation}
Notice that the Poisson brackets are replaced by the commutators coherently with the rule $\{\cdot, \cdot\} \to  [\cdot, \cdot]/\imath \hbar$. In the given realisation, the Casimir operator reads: \begin{equation}
\hat{C}^{[1]}=(\hat{J}_3^{[1]})^2-\frac{1}{2}\{\hat{J}_+^{[1]}, \hat{J}_-^{[1]}\}=\frac{1}{16}(3 \hbar^2-4 a_1)\mathds{1}=\hbar^2 \textsf{k}_1(1-\textsf{k}_1) \mathds{1} \, ,
\label{cas1}
\end{equation}
where we introduced the quantity $a_1:=\hbar^2\bigl((2\textsf{k}_1-1)^2-1/4\bigl)=\hbar^2(4\textsf{k}_1-1)(4\textsf{k}_1-3)/4$.

In total analogy with the classical case, the $n$D realisation can be obtained by rising the dimensionality with the help of the coproduct. Notice that in higher dimensions we deal with the $2n$ basis operators $(\hbx, \hbp)$, the latter defining  the Heisenberg algebra $\mathfrak{h}_n$: 
\begin{equation}
[\hat{x}_i, \hat{x}_j]=[\hat{p}_i, \hat{p}_j]=0  \, ,\quad [\hat{x}_i, \hat{p}_j]=\imath \hbar \delta_{ij} \mathds{1} \, ,
\label{eq:Heisbq}
\end{equation}
the quantum analog of \eqref{eq:Heis}.
So, in general, the quantum integrals in $n$D will be well-defined Hermitian operators in the UEA of the Heisenberg algebra $\mathfrak{h}_n$ (or convergent series in the basis $(\hbx, \hbp)$) \cite{MillerPostWinternitz2013R}.
Now, by applying the coproduct \eqref{ncopr} on the basis generators and considering the new realisation in terms of differential operators we get the $n$D extension:
\begin{align}
\hat{J}_+^{[n]} = \frac{1}{2}\biggl(-\hbar^2\Delta+\sum_{j=1}^n\frac{a_j}{x_j^2} \biggl)\qquad \hat{J}_-^{[n]} =\frac{1}{2}\bx^2\qquad \hat{J}_3^{[n]}= -\frac{\imath \hbar}{2} (\bx \cdot \nabla+n/2) \, ,
\end{align}
which represents the quantum analog of \eqref{eqngen}. Here $\Delta = \nabla \cdot \nabla = \sum_{j=1}^n \partial_{x_j}^2$ is the Laplacian, $\nabla=(\partial_{x_1},\dots,\partial_{x_n})$ being the vector of partial derivatives. The above operators close to give\footnote{Abstractly, this result can be deduced directly from the homomorphism property of $\Delta^{[n]}$ \cite{0305-4470-31-16-009}.}:
\begin{equation}
[\hat{J}_-^{[n]}, \hat{J}_+^{[n]}]= 2\imath \hbar \hat{J}_3^{[n]} \quad [\hat{J}_3^{[n]}, \hat{J}_{\pm}^{[n]}]=\pm  \imath \hbar  \hat{J}_\pm^{[n]} \, .
\label{eq:quantumgenn}
\end{equation}
At this point, we observe that the quantum analog of the Hamiltonian \eqref{eq:HJ} arises as:
\begin{equation}
\hat{H}=\hat{J}_+^{[n]}  \, + \, V\bigl(\,(2 \hat{J}_-^{[n]})^{1/2}\,\bigl) = \frac{1}{2}\biggl(\hbp^2+\sum_{j=1}^n \frac{a_j}{\hat{x}_j^2}\biggl)\,+\,V(|\hbx|) = \frac{1}{2}\biggl(-\hbar^2\Delta+\sum_{j=1}^n \frac{a_j}{x_j^2}\biggl)\,+\,V(|\bx|)  \, ,
\label{eq:Hq}
\end{equation}
and the total Casimir  operator $\hat{C}^{[n]}$, in the given  realisation, reads:
\begin{align}
\hat{C}^{[n]}= (\hat{J}_3^{[n]})^2-\frac{1}{2}\{\hat{J}_+^{[n]}, \hat{J}_-^{[n]}\}&=-\frac{1}{4}\biggl(\sum_{1 \leq i< j}^n \bigl(\hat{L}_{ij}^2+a_i \frac{\hat{x}_j^2}{\hx_i^2}+a_j \frac{\hx_i^2}{\hx_j^2}\bigl)+\sum_{i=1}^n a_i+\frac{\hbar^2}{4}n(n-4)\biggl) \nonumber \\
&=-\frac{1}{4}\biggl(\sum_{1 \leq i< j}^n \hat{L}_{ij}^2+\hbx^2 \sum_{j=1}^n \frac{a_j}{\hx_j^2}+\frac{\hbar^2}{4}n(n-4)\biggl) \nonumber \\
&=\sum_{1 \leq i< j}^n \hat{C}_{ij}-(n-2)\sum_{i=1}^n \hat{C}_{i} \nonumber \\
&= \sum_{1 \leq i< j}^n \hat{P}_{ij}+\sum_{i=1}^n \hat{C}_{i} \, .
\end{align}
where we introduced the new operators (the quantum analog of \eqref{cijc}):
\begin{equation}
\hat{C}_{ij}:=-\frac{1}{4}\biggl(\hat{L}_{ij}^2+a_i \frac{\hx_j^2}{\hx_i^2}+a_j \frac{\hx_i^2}{\hx_j^2} +a_i+a_j-\hbar^2\biggl) \, ,\qquad \hat{C}_{i}:= \frac{1}{16}(3 \hbar^2-4 a_i) \, , \qquad \hat{P}_{ij}:=\hat{C}_{ij}-\hat{C}_i-\hat{C}_j
\label{cij}
\end{equation}
with $\hat{L}_{ij} =\hx_i \hp_j-\hx_j \hp_i = - \imath \hbar x_i \partial_{x_j} +\imath \hbar  x_j \partial_{x_i}=-\hat{L}_{ji}$. Thus, the left and right Casimirs at the quantum level read:
\begin{equation}
\hat{C}^{[m]}=\sum_{1 \leq i< j}^m \hat{P}_{ij}+\sum_{i=1}^m \hat{C}_i \qquad \hat{C}_{[m]}=\sum_{n-m+1 \leq i< j}^n \hat{P}_{ij}+\sum_{i=n-m+1}^n \hat{C}_i  \qquad (m=1, \dots, n) 
\label{qCas}
\end{equation}
and, for $m=n$, they coincide to give the total Casimir $\Delta^{[n]}(C)$, i.e.:

\begin{equation}
\hat{C}^{[n]}=\hat{C}_{[n]}=\sum_{1 \leq i< j}^n \hat{P}_{ij}+\sum_{i=1}^n \hat{C}_i \, .
\label{eq:totcas}
\end{equation} 
In complete analogy to the classical case, it arises as a linear combination of the $\binom{n}{2}$ operators $\hat{P}_{ij}$ and the $n$ central elements $\hat{C}_i$, now realised as in \eqref{cij}.
Moreover, as a consequence of the underlying coalgebra symmetry the set $U  = U_1\, \cup \,U_2$ with  $U_1 := \{\hat{C}_{[m]}\}$, $U_2:=\{\hat{C}^{[m]}\}$ for $m=1, \dots, n$ provides $2n-3$  algebraically independent quantum integrals and, since the elements in each subset commute each other, they define two abelian subalgebras composed by $n-1$ commuting elements, just like the classical case. So, the quantum Hamiltonian \eqref{eq:Hq} is QMS for any choice of the function $V=V(|\hbx|)$. Furthermore, each of the two sets $V_i :=U_i \cup \{\hat{H}\}$ ($i=1,2$) is composed by $n$ algebraically independent commuting quantum integrals.

Again, the connection with the Racah algebra $R(n)$ is made explicit by considering the \textquotedblleft quantum generators\textquotedblright:
\begin{equation}
\hat{P}_{ij}=-\frac{1}{4}\hat{Q}_{ij}\, ,
\label{genqRacah}
\end{equation}
where $\hat{Q}_{ij}= \hat{L}_{ij}^2+a_i \dfrac{\hx_j^2}{\hx_i^2}+a_j \dfrac{\hx_i^2}{\hx_j^2}+\dfrac{\hbar^2}{2}$, together with the quantum analog of \eqref{eq:f}, i.e. the three indices quantum integrals:
\begin{equation}
\hat{F}_{ijk}:=\frac{1}{2 \imath \hbar} [\hat{P}_{ij}, \hat{P}_{jk}] \qquad (i \neq j \neq k) \, .
\label{eq:qf}
\end{equation}
With the above definitions, for $i,j,k,l,m,r \in \{1,\dots, n\}$ all different, we have:
\begin{equation}
[\hat{P}_{ij}, \hat{H}]=[\hat{F}_{ijk},\hat{H}] =0\, , \qquad  [\hat{P}_{ij}, \hat{P}_{kl}]=0 \, , \qquad [\hat{P}_{ij}, \hat{P}_{ik}+\hat{P}_{jk}]=0
\end{equation}
and we get a quadratic algebra whose defining relations are those of the Racah algebra $R(n)$:
\begin{align}
&[\hat{P}_{ij},\hat{P}_{jk}]=: 2 \imath \hbar \hat{F}_{ijk} \hskip 5.2cm (\hat{F}_{ijk}=-{\hat{F}_{jik}}=-\hat{F}_{ikj}) \label{qracah1}\\
&[\hat{P}_{jk}, \hat{F}_{ijk}]=\imath \hbar (\hat{P}_{ik}\hat{P}_{jk}-\hat{P}_{jk}\hat{P}_{ij}+2\hat{P}_{ik}\hat{C}_j-2\hat{P}_{ij}\hat{C}_k)\\
&[\hat{P}_{kl}, \hat{F}_{ijk}]=\imath \hbar(\hat{P}_{ik}\hat{P}_{jl}-\hat{P}_{il}\hat{P}_{jk})\\
&[\hat{F}_{ijk}, \hat{F}_{jkl}]=\imath \hbar(\hat{F}_{jkl}\hat{P}_{ij}-\hat{F}_{ikl}(\hat{P}_{jk}+2 \hat{C}_j)-\hat{F}_{ijk}\hat{P}_{jl})\\
&[\hat{F}_{ijk}, \hat{F}_{klm}]=\imath \hbar(\hat{F}_{ilm} \hat{P}_{jk}-\hat{P}_{ik} \hat{F}_{jlm})
\label{qracah5}
\end{align}
together with $[\hat{F}_{ijk},\hat{F}_{lmr}]=0$. This means that the Euclidean quantum Hamiltonian \eqref{eq:Hq}, for any suitable choice of the potential function $V=V(|\hbx|)$, turns out to be endowed with quantum integrals that close to give the quadratic  algebra $R(n)$. The left and right  Casimir invariants, in the given realisation appear as specific linear combinations of them, as shown in \eqref{qCas}. Notice that the classical relations \eqref{eq:PoissonRacah1}-\eqref{eq:PoissonRacah5} are obtained in the limit $\hbar \to 0$ by replacing the commutator $[\hat{A}, \hat{B}]/(\imath \hbar)=\hat{C}$ with the Poisson bracket $\{A, B\}=C$. 

Now, as for the classical case, it is possible to show that the closure of the quadratic algebra is ensured by the high-order relations among the quantum generators, now expressed in terms of their symmetric products, i.e:
\begin{align}
&\hat{F}_{ijk}^2-\hat{C}_i \hat{P}_{jk}^2-\hat{C}_j \hat{P}_{ik}^2-\hat{C}_k \hat{P}_{ij}^2+\frac{1}{6}\{\hat{P}_{ij},\hat{P}_{jk},\hat{P}_{ik}\}+4 \hat{C}_i \hat{C}_j \hat{C}_k \nonumber\\
&\hskip 0.65cm+\frac{\hbar^2}{3}\bigl(\{\hat{P}_{ij},\hat{P}_{jk}\}+\{\hat{P}_{ij},\hat{P}_{ik}\}+\{\hat{P}_{ik},\hat{P}_{jk}\}+2 \hat{C}_i \hat{P}_{jk}+2\hat{C}_j \hat{P}_{ik}+2\hat{C}_k \hat{P}_{ij}\bigl)=0\\
&\{\hat{F}_{ijk},\hat{F}_{jkl}\}-\frac{1}{6}(\{\hat{P}_{il}, \hat{P}_{jk},\hat{P}_{jk}\}-\{\hat{P}_{ij},\hat{P}_{jk},\hat{P}_{kl}\}-\{\hat{P}_{ik}, \hat{P}_{jk},\hat{P}_{jl}\})-\hat{C}_j\{\hat{P}_{ik},\hat{P}_{kl}\}-\hat{C}_k\{\hat{P}_{ij},\hat{P}_{jl}\}+4 \hat{C}_j\hat{C}_k\hat{P}_{il} \nonumber\\
&\hskip 1.8cm+\frac{\hbar^2}{3}\bigl(\{\hat{P}_{ij},\hat{P}_{kl}\}+\{\hat{P}_{ik},\hat{P}_{jl}\}+\{\hat{P}_{il},\hat{P}_{jk}\}\bigl)=0\\
&\{\hat{F}_{ijk},\hat{F}_{klm}\}-\frac{1}{6}(\{\hat{P}_{il}, \hat{P}_{jk},\hat{P}_{km}\}+\{\hat{P}_{ik},\hat{P}_{jm},\hat{P}_{kl}\}-\{\hat{P}_{im}, \hat{P}_{jk},\hat{P}_{kl}\}-\{\hat{P}_{ik}, \hat{P}_{jl},\hat{P}_{km}\}) \nonumber \\ 
&\hskip 1.9cm-\hat{C}_k\{\hat{P}_{im},\hat{P}_{jl}\}+\hat{C}_k\{\hat{P}_{il},\hat{P}_{jm}\} =0\\
&\{\hat{F}_{ijk}, \hat{F}_{lmr}\}-\frac{1}{6}(\{\hat{P}_{il}, \hat{P}_{jr}, \hat{P}_{km}\}+\{\hat{P}_{ir}, \hat{P}_{jm}, \hat{P}_{kl}\}+\{\hat{P}_{im},\hat{P}_{jl},\hat{P}_{kr}\} \nonumber \\
&\hskip 2.25cm-\{\hat{P}_{im}, \hat{P}_{jr}, \hat{P}_{kl}\}-\{\hat{P}_{ir}, \hat{P}_{jl},\hat{P}_{km}\}-\{\hat{P}_{il},\hat{P}_{jm},\hat{P}_{kr}\}) =0
\end{align}
where $\{\hat{A}, \hat{B}, \hat{C}\}:=\hat{A}\hat{B}\hat{C}+\hat{A}\hat{C}\hat{B}+\hat{B}\hat{A}\hat{C}+\hat{B}\hat{C}\hat{A}+\hat{C}\hat{A}\hat{B}+\hat{C}\hat{B}\hat{A}$ is the symmetrizer of three operators. We notice that some terms, as for example $\{\hat{P}_{ij}, \hat{P}_{kl}\}$ can be alternatively written as  $2\hat{P}_{ij} \hat{P}_{kl}$ (this is because $[\hat{P}_{ij}, \hat{P}_{kl}]=0$). Also, we remark that the above relations collapse to the classical ones \eqref{eq:ho1}-\eqref{eq:ho4} in the classical $\hbar \to 0$ limit,  taking into account that $\frac{1}{6}\{\hat{A},\hat{B},\hat{C}\} \to ABC$ and $\frac{1}{2}\{\hat{A},\hat{B}\} \to AB$.

Such quantum closure relations had only been observed previously in the context of two and three dimensional models \cite{Kalnins_2007, 2011SIGMA...7..051K}. Our results show that it is a general feature that such higher order relations appear in quadratic algebras in classical as well as in quantum mechanics. They are similar to the Serre-type relations appearing in Lie algebras and provide constraints useful in constructing representation of the quadratic algebra.

\section{Maximal superintegrability and embedding of the Racah algebra $\boldsymbol{R(n)}$}
\label{sec3}

 So far we have shown how it is possible to realise the generalized Racah algebra $R(n)$ in terms of suitable symplectic or differential realisations and that its classical and quantum generators are conserved quantities for the QMS Hamiltonians \eqref{eq:HJ} and \eqref{eq:Hq} respectively. In this section, we specialise the functional form of the potential $V=V(r)$ in such a way to consider two well-known MS Euclidean models appearing as particular subcases of the $n$D family,  namely the $n$D Smorodinsky-Winternitz and the $n$D generalized Kepler-Coulomb system.  We show that the Racah algebra $R(n)$ will appear as embedded inside their full symmetry algebra.

\subsection{The nD Smorodinsky-Winternitz system}
\label{SW}

The Smorodinsky-Winternitz (SW) system \cite{PhysRevA.41.5666, FRIS1965354, Makarov1967, EVANS1990483, doi:10.1063/1.529449} is a well-known second-order superintegrable model. In $n$ dimensions, it arises as a specific subcase of \eqref{eq:HJ} and \eqref{eq:Hq} with classical and quantum potential functions:

\begin{equation}
V=V(|\bx|)=\frac{1}{2} \omega^2 \bx^2 \, , \qquad V=V(|\hbx|)=\frac{1}{2} \omega^2 \hbx^2 \, ,
\label{eq:SW}
\end{equation}
where $\omega \in \mathbb{R}$ is a real parameter. In terms of the generators $J_{\pm, 3}$ it can be thus expressed as follows:

\begin{equation}
H = J_+^{[n]}+\omega^2 J_-^{[n]} .
\label{eq:SWmodel}
\end{equation}
In what follows, our objective is to construct the full symmetry algebra of this MS system both in classical and quantum mechanics. The main aim is to show how the Racah algebra $R(n)$ appears as embedded inside a larger algebra which includes $n$ additional constants of motion/quantum integrals arising for the specific choice \eqref{eq:SW}.

\label{subsec3c}
\subsubsection{The nD classical Smorodinky-Winternitz system}

 Let us consider the classical SW Hamiltonian:
\begin{equation}
H = \frac{1}{2}\biggl(\bp^2+\sum_{j=1}^n \frac{a_j}{x_j^2}+\omega^2 \bx^2 \biggl)\, .
\label{eq:cSW}
\end{equation}
Besides the $\binom{n}{2}$ common constants $Q_{ij}$ given in \eqref{eq:Qij} (or the equivalent $P_{ij}$), this Hamiltonian is endowed with the following  $n$ additional second-order integrals, the 1D Hamiltonians:
\begin{equation}
H_i=\frac{1}{2}\biggl(p_i^2+ \frac{a_i}{x_i^2}+\omega^2 x_i^2 \biggl)\qquad (i = 1, \dots, n) 
\label{addint}
\end{equation}
such as $H= \sum_{i=1}^n H_i$. Let us introduce the  new two indices classical generators: \begin{equation}
G_{ij} :=\frac{1}{2}\{H_i, P_{ij}\}  \qquad (i \neq j) \, .
\label{Gij}
\end{equation}
Then, for $i,j,k,l,m \in \{1,\dots, n\}$ all different, we have: 
\begin{equation}
\{H_i, H\}=\{P_{ij}, H\}=\{G_{ij}, H\}=0 \, , \qquad \{H_i, H_j\}= \{H_i, P_{jk}\}=0 \, ,
\label{addrel}
\end{equation}
and, together with the defining relations of the Racah algebra $R(n)$ given in \eqref{eq:PoissonRacah1}-\eqref{eq:PoissonRacah5},  we close a quadratic Poisson algebra whose additional defining relations are:
\begin{align}
&\{H_{i},P_{ij}\}=:2 G_{ij}\hskip 5.5cm (G_{ji}=-G_{ij}) \\
&\{H_i, G_{ij}\}=-H_i H_j-2\omega^2 P_{ij}\\
&\{H_i, G_{jk}\}=0\\
&\{P_{ij},G_{ij}\}=H_j (P_{ij}+2C_i)-H_i (P_{ij}+2C_j)\\
&\{P_{ij}, G_{ik}\}=H_j P_{ik}-H_i P_{jk}\\
&\{P_{ij},G_{kl}\}=0\\
&\{H_i, F_{ijk}\}=H_k P_{ij}-H_j P_{ik}\\
&\{G_{ij}, G_{ik}\}=H_i G_{jk}\\
&\{G_{ij},G_{kl}\}=0\\
&\{G_{ij}, F_{ijk}\}=-P_{ij}(G_{ik}+ G_{jk})\\
&\{G_{ij},F_{jkl}\}=P_{jk}G_{il}-P_{jl}G_{ik}\\
&\{G_{ij},F_{klm}\}=0 \, ,
\end{align}
Of course, all the generators cannot be functionally independent. However, due to the underlying coalgebra symmetry of the model, it is possible to extrapolate a set of $2n-1$ functionally independent constants. As a matter of fact, each of the $n$ sets $V_i =\{C_{[m]}\} \cup \{C^{[m]}\}  \cup \{H\} \cup \{H_i\}$, at a fixed index $i =1, \dots, n$, turns out to be composed by $2n-1$ functionally independent constants of motion, $2n-3$ of them given by the left and right Casimir invariants of the coalgebra (including the total Casimir $C_{[n]}=C^{[n]}$) \cite{Ballesteros_2004}.

Once again, it is possible to obtain additional high-order relations among the above classical generators:
\begin{align}
&G_{ij}^2-C_j H_i^2-C_i H_j^2+P_{ij}H_i H_j+\omega^2(P_{ij}^2-4 C_i C_j) =0\label{choSW1}\\
& 2G_{ij}G_{jk}-P_{ij}H_jH_k-P_{jk}H_iH_j+P_{ik}H_j^2+2 C_j H_i H_k  -2 \omega^2 (P_{ij}P_{jk}-2 C_j P_{ik})=0\\
&2 G_{ij}G_{kl}-P_{ik}H_jH_l-P_{jl}H_iH_k+P_{jk}H_iH_l+P_{il}H_jH_k-2 \omega^2(P_{jl}P_{ik}-P_{il}P_{jk} )=0\\
& 2 G_{ij} F_{ijk}-H_k P_{ij}^2+H_j P_{ij}P_{ik}+H_iP_{ij}P_{jk}-2C_iH_jP_{jk}-2C_jH_iP_{ik}+4C_iC_jH_k=0\\
&2 G_{ij}F_{jkl}- H_k P_{ij}P_{jl}-H_j P_{il}P_{jk}+H_j P_{jl}P_{ik}+H_l P_{ij}P_{jk}-2 C_j H_{l}P_{ik}+2 C_jH_kP_{il}=0\\
&2 G_{ij}F_{klm}-H_k P_{im}P_{jl}-H_mP_{il} P_{jk}- H_{l}P_{jm} P_{ik}+ H_kP_{il}P_{jm}+H_l P_{im}P_{jk}+H_m P_{jl}P_{ik}=0 \label{choSW6}\, ,
\end{align}
together with the ones we listed before in \eqref{eq:ho1}-\eqref{eq:ho4}.  Note again that the Racah subalgebra generators as well as the extra integrals are all involved in the cubic constraints.

\subsubsection{The nD quantum Smorodinksy-Winternitz system}
\label{subsec3q}

 Let us consider the quantum SW Hamiltonian:
\begin{equation}
\hat{H} =\frac{1}{2}\biggl(\hbp^2+\sum_{j=1}^n \frac{a_j}{\hat{x}_j^2}+ \omega^2 \hbx^2 \biggl)  =  \frac{1}{2}\biggl(-\hbar^2 \Delta+\sum_{j=1}^n \frac{a_j}{x_j^2}+\omega^2 \bx^2 \biggl)\, .
\label{eq:qSW}
\end{equation}

 Besides the quantum integrals $\hat{Q}_{ij}$ (or $\hat{P}_{ij}$) given in \eqref{genqRacah}, as for the classical case this system is endowed with the following  additional $n$ quantum integrals of the second-order, the 1D  Hamiltonians:
\begin{equation}
\hat{H}_i=\frac{1}{2}\biggl(\hp_i^2+\frac{a_i}{\hat{x}_i^2},+\,  \omega^2 \hat{x}_i^2 \biggl)  = \frac{1}{2}\biggl(-\hbar^2\partial_{x_i}^2+ \frac{a_i}{x_i^2}+\omega^2 x_i^2 \biggl) \qquad (i = 1, \dots, n)
\label{qSW}
\end{equation}
such as $\hat{H}= \sum_{i=1}^n \hat{H}_i$. If we introduce the  new two indices quantum integrals: \begin{equation}
\hat{G}_{ij} :=\frac{1}{2 \imath \hbar}[\hat{H}_i, \hat{P}_{ij}]  \qquad (i \neq j)
\label{Gqij}
\end{equation}
then, for $i,j,k,l,m \in \{1,\dots, n\}$ all different, we get: 
\begin{equation}
[\hat{H}_i, \hat{H}]=[\hat{P}_{ij}, \hat{H}]=[\hat{G}_{ij},\hat{H}]=0 \, , \qquad [\hat{H}_i, \hat{H}_j]= [\hat{H}_i, \hat{P}_{jk}]=0 \, ,
\label{eq:qrel}
\end{equation}
and, together with the defining relations of the Racah algebra $R(n)$  given in \eqref{qracah1}-\eqref{qracah5},  we close a quadratic algebra whose additional defining relations are:
\begin{align*}
&[\hat{H}_{i}, \hat{P}_{ij}]=:2 \imath \hbar \hat{G}_{ij}\hskip 5.5cm (\hat{G}_{ji}=-\hat{G}_{ij}) \\
&[\hat{H}_i, \hat{G}_{ij}]=-\imath \hbar(\hat{H}_i \hat{H}_j+2\omega^2 \hat{P}_{ij})\\
&[\hat{H}_i, \hat{G}_{jk}]=0\\
&[\hat{P}_{ij},\hat{G}_{ij}]=\frac{\imath \hbar}{2}(\{\hat{H}_j, \hat{P}_{ij}\}-\{\hat{H}_i, \hat{P}_{ij}\}+4 \hat{C}_i \hat{H}_j-4 \hat{C}_j \hat{H}_i)\\
&[\hat{P}_{ij}, \hat{G}_{ik}]=\imath \hbar(\hat{H}_j \hat{P}_{ik}-\hat{H}_i \hat{P}_{jk})\\
&[\hat{P}_{ij}, \hat{G}_{kl}]=0\\
&[\hat{H}_i, \hat{F}_{ijk}]=\imath \hbar(\hat{H}_k \hat{P}_{ij}-\hat{H}_j \hat{P}_{ik})\\
&[\hat{G}_{ij}, \hat{G}_{ik}]=\imath \hbar \hat{H}_i \hat{G}_{jk}\\
&[\hat{G}_{ij}, \hat{G}_{kl}]=0\\
&[\hat{G}_{ij}, \hat{F}_{ijk}]=- \frac{\imath\hbar}{2}(\{\hat{P}_{ij}, \hat{G}_{ik}\}+\{\hat{P}_{ij}, \hat{G}_{jk}\})\\
&[\hat{G}_{ij}, \hat{F}_{jkl}]=\frac{\imath \hbar}{2}(\{\hat{P}_{jk}, \hat{G}_{il}\}-\{\hat{P}_{jl}, \hat{G}_{ik}\})\\
&[\hat{G}_{ij}, \hat{F}_{klm}]=0
\end{align*}

 In complete analogy with the classical case, all the generators cannot be algebraically independent. However, due to the coalgebra symmetry of the model, it is possible to extrapolate the same set of $2n-1$ algebraically independent quantum integrals. Once again, each of the $n$ sets $V_i =\{\hat{C}_{[m]}\} \cup \{\hat{C}^{[m]}\}  \cup \{\hat{H}\} \cup \{\hat{H}_i\}$, at a fixed index $i =1, \dots, n$, turns out to be composed by $2n-1$ algebraically independent quantum integrals, $2n-3$ of them given by the left and right Casimirs of the coalgebra.  Again, the closure of the algebra is ensured by the following additional high-order relations among the above quantum generators:
 
\begin{align}
&\hat{G}_{ij}^2-\hat{C}_j \hat{H}_i^2-\hat{C}_i \hat{H}_j^2+\frac{1}{6}\{\hat{P}_{ij}, \hat{H}_i, \hat{H}_j\}+\omega^2(\hat{P}_{ij}^2-4 \hat{C}_i \hat{C}_j)+\frac{\hbar^2}{3}(\{\hat{H}_i,\hat{H}_j\}-2\omega^2 \hat{P}_{ij})=0\\
&\{\hat{G}_{ij}, \hat{G}_{jk}\}-\frac{1}{6}(\{\hat{P}_{ij},\hat{H}_j,\hat{H}_k\}+\{\hat{P}_{jk},\hat{H}_i,\hat{H}_j\}-\{\hat{P}_{ik},\hat{H}_j,\hat{H}_j\})+2 \hat{C}_j\hat{H}_i \hat{H}_k  -\omega^2 \bigl(\{\hat{P}_{ij},\hat{P}_{jk}\}-4 \hat{C}_j \hat{P}_{ik}\bigl)=0\\
&\{\hat{G}_{ij},\hat{G}_{kl}\}-\frac{1}{6}(\{\hat{P}_{ik},\hat{H}_j,\hat{H}_l\}+\{\hat{P}_{jl}, \hat{H}_i,\hat{H}_k\}-\{\hat{P}_{jk}, \hat{H}_i,\hat{H}_l\}-\{\hat{P}_{il}, \hat{H}_j, \hat{H}_k\}) - \omega^2 \bigl(\{\hat{P}_{ik},\hat{P}_{jl}\}- \{\hat{P}_{il}, \hat{P}_{jk}\}\bigl)=0\\
&\{\hat{G}_{ij},\hat{F}_{ijk}\}-\frac{1}{6}(\{\hat{H}_k, \hat{P}_{ij},\hat{P}_{ij}\}-\{\hat{H}_j, \hat{P}_{ij},\hat{P}_{ik}\}-\{\hat{H}_i,\hat{P}_{ij},\hat{P}_{jk}\})-\hat{C}_i\{\hat{H}_j, \hat{P}_{jk}\}-\hat{C}_j\{\hat{H}_i,\hat{P}_{ik}\}+4\hat{C}_i\hat{C}_j\hat{H}_k \nonumber\\
&\hskip 1.7cm+\frac{\hbar^2}{3}(\{\hat{H}_i, \hat{P}_{jk}\}+\{\hat{H}_{j},\hat{P}_{ik}\}+\{\hat{H}_{k},\hat{P}_{ij}\})=0\\
&\{\hat{G}_{ij}, \hat{F}_{jkl}\}-\frac{1}{6}(\{\hat{H}_j, \hat{P}_{il}, \hat{P}_{jk}\}+\{\hat{H}_k, \hat{P}_{ij}, \hat{P}_{jl}\}-\{\hat{H}_j, \hat{P}_{jl}, \hat{P}_{ik}\}-\{\hat{H}_l, \hat{P}_{ij}, \hat{P}_{jk}\})- \hat{C}_j \{\hat{H}_{l},\hat{P}_{ik}\}+ \hat{C}_j\{\hat{H}_k, \hat{P}_{il}\}=0\\
&\{\hat{G}_{ij}, \hat{F}_{klm}\}-\frac{1}{6}(\{\hat{H}_k, \hat{P}_{im}, \hat{P}_{jl}\}+ \{\hat{H}_{l}, \hat{P}_{jm}, \hat{P}_{ik}\}+\{\hat{H}_m, \hat{P}_{il}, \hat{P}_{jk}\} \nonumber\\
&\hskip 2.2cm- \{\hat{H}_k, \hat{P}_{il}, \hat{P}_{jm}\}-\{\hat{H}_l, \hat{P}_{im}, \hat{P}_{jk}\}-\{\hat{H}_m, \hat{P}_{jl}, \hat{P}_{ik}\}) =0\, .
\end{align}

The above high-order relations reduce to \eqref{choSW1}-\eqref{choSW6} in the classical $\hbar \to 0$ limit.  These relations are reminiscent of the Serre-type relations in Lie algebras and are useful in constructing representations of the quadratic symmetry algebra.

\subsection{The nD generalized Kepler-Coulomb system}
\label{gKCc}

 The generalized  Kepler-Coulomb (gKC) system \cite{doi:10.1063/1.2840465,1751-8121-42-24-245203, 2011SIGMA...7..054T} is a well-known fourth-order superintegrable model. In $n$ dimensions, it arises as a specific subcase of \eqref{eq:HJ} and \eqref{eq:Hq} with classical and quantum potential functions:
\begin{equation}
V=V(|\bx|)=-\mu/|\bx|\, , \qquad V=V(|\hbx|)=-\mu/|\hbx|\, ,
\label{eq:gKC}
\end{equation}
where $\mu \in \mathbb{R}$ is a real parameter. In terms of the generators $J_{\pm,3}$ it can be thus expressed as:
\begin{equation}
H = J_+^{[n]}-\mu\bigl( 2J_-^{[n]}\bigl)^{-1/2} .
\label{eq:gKCmodel}
\end{equation}

\subsubsection{The nD classical generalized Kepler-Coulomb system }
\label{subsec4c}

 Let us consider the classical generalized Kepler-Coulomb Hamiltonian:
\begin{equation}
H = \frac{1}{2}\biggl(\bp^2+\sum_{j=1}^n \frac{a_j}{x_j^2}\biggl)-\frac{\mu}{|\bx|}  \, .
\label{eq:classint}
\end{equation}
Besides the $\binom{n}{2}$ common constants
$Q_{ij}$ (or $P_{ij}$), this system turns out to be endowed with the following $n$ additional fourth-order integrals:
\begin{equation}
R_i=\biggl[\sum_{j=1}^{n} \biggl(L_{ij}  p_j+x_i \frac{a_j}{x_j^2}\biggl)-x_i \frac{\mu}{|\bx|} \biggl]^2+\frac{a_i}{x_i^2}(\bx\cdot \bp)^2 \qquad (i=1, \dots, n) ,
\label{gc}
\end{equation}
as it can be checked by direct computations. Notice that the following relation holds true:
\begin{equation}
\sum_{i=1}^n R_i = -8 H C^{[n]}+\mu^2 \, ,
\label{funrel}
\end{equation}
where $C^{[n]}$ is the total Casimir \eqref{eq:total}. This formula boils down to the usual functional relation holding for the constants of motion of the KC problem on the $n$D Euclidean space when $a_j=0 \quad \forall\, j=1, \dots, n$. In this case, in fact, the constants \eqref{gc} define the vector $\boldsymbol{R}=(A_1^2, \dots, A_n^2)$, where $A_i = \sum_{j=1}^{n} L_{ij}  p_j-x_i\frac{k}{|\bx|}$ are the components of the $n$D LRL vector $\boldsymbol{A}$.  Moreover, the total Casimir reads:
\begin{equation}
C^{[n]}=-\frac{1}{4}\sum_{1 \leq i<j}^n L_{ij}^2=-\frac{1}{4}\boldsymbol{L}^2
\label{totcasrot}
\end{equation}
so that \eqref{funrel} results in:
\begin{equation}
\sum_{i=1}^n A_i^2 = 2 H_{\textsf{KC}} \boldsymbol{L}^2+\mu^2 \, ,  \qquad \biggl(H_{\textsf{KC}}=\frac{\bp^2}{2}- \frac{\mu}{|\bx|}\biggl) \, .
\label{eq:funcrelKC}
\end{equation}

At this point, in complete analogy with construction holding for the SW case,  if we introduce the  new  two indices classical generators: \begin{equation}
G_{ij} :=\frac{1}{2}\{R_i, P_{ij}\}  \qquad (i \neq j)
\label{GKCij}
\end{equation}
then, for $i,j,k,l,m \in \{1,\dots, n\}$ all different, we get: 
\begin{equation}
\{R_i, H\}=\{P_{ij}, H\}=\{G_{ij}, H\}=0 \, , \qquad \{R_i, P_{jk}\}=0 \, ,
\label{addrel}
\end{equation}
and, together with the defining relations of the Racah algebra $R(n)$ given in \eqref{eq:PoissonRacah1}-\eqref{eq:PoissonRacah5},  we close a quadratic Poisson algebra whose additional defining relations are:

\begin{align}
&\{R_{i},P_{ij}\}=:2 G_{ij}\hskip 5.5cm (G_{ji}=-G_{ij}) \\
&\{R_i,R_j\}=-16 H G_{ij} \\
&\{R_i, G_{ij}\}=-R_i R_j+8 H (R_i P_{ij}-2C_iR_j)\\
&\{R_i, G_{jk}\}=8 H (P_{ik}R_j- P_{ij}R_k)\\
&\{P_{ij},G_{ij}\}=R_j(P_{ij}+2C_i)-R_i(P_{ij}+2C_j)\\
&\{P_{ij}, G_{ik}\}=R_j P_{ik}-R_i P_{jk}\\
&\{P_{ij},G_{kl}\}=0\\
&\{R_i, F_{ijk}\}=R_k P_{ij}-R_j P_{ik}\\
&\{R_i, F_{jkl}\}=0\\
&\{G_{ij}, G_{ik}\}=G_{jk}R_{i}+8HF_{ijk}R_i \\
&\{G_{ij},G_{kl}\}=8H(R_j F_{ikl}-R_i F_{jkl})\\
&\{G_{ij}, F_{ijk}\}=-P_{ij}(G_{ik}+G_{jk})\\
&\{G_{ij},F_{jkl}\}=P_{jk}G_{il}-P_{jl}G_{ik}\\
&\{G_{ij},F_{klm}\}=0 \, .
\end{align}
Also in this case, all the generators cannot be functionally independent. However, due to the coalgebra symmetry of the model, it is possible to extrapolate a set of $2n-1$ functionally independent constants. In this case, each of the $n$ sets $V_i =\{C_{[m]}\} \cup \{C^{[m]}\}  \cup \{H\} \cup \{R_i\}$, at a fixed index $i =1, \dots, n$, turns out to be composed by $2n-1$ functionally independent constants of motion, $2n-3$ of them given by the usual left and right Casimirs of the coalgebra together with the total Casimir $C_{[n]}=C^{[n]}$ \cite{1751-8121-42-24-245203}.  Let us remark that even with the presence of integrals $R_{i}$ of order 4 and $G_{ij}$ of order 5, the symmetry algebra is still quadratic,  which is quite remarkable.

Also for this model, high-order relations among the generators can be found, they read:
\begin{align}
&G_{ij}^2-C_i R_j^2-C_j R_i^2+P_{ij}R_i R_j=0 \label{hocKC1} \\
& 2G_{ij}G_{jk}-P_{ij}R_jR_k-P_{jk}R_iR_j+P_{ik}R_j^2+2 C_j R_i R_k=0 \\
&2 G_{ij}G_{kl}-P_{ik}R_jR_l-P_{jl}R_iR_k+P_{jk}R_iR_l+P_{il}R_jR_k=0\\
& 2 G_{ij} F_{ijk}-R_k P_{ij}^2+R_jP_{ij}P_{ik}+R_iP_{ij}P_{jk}-2C_iR_jP_{jk}-2C_jR_iP_{ik}+4C_iC_jR_k=0\\
&2 G_{ij}F_{jkl}- R_k P_{ij}P_{jl}-R_j P_{il}P_{jk}+R_j P_{jl}P_{ik}+R_l P_{ij}P_{jk}-2 C_j R_{l}P_{ik}+2 C_jR_kP_{il}=0\\
&2 G_{ij}F_{klm}-R_k P_{im}P_{jl}-R_mP_{il} P_{jk}- R_{l}P_{jm} P_{ik}+ R_kP_{il}P_{jm}+R_l P_{im}P_{jk}+R_m P_{jl}P_{ik}=0 \, .
\label{hocKC6}
\end{align}

Here, these relations are cubic in terms of the generators but are of higher order in terms of the momenta when using explicit realizations.

\subsubsection{The nD quantum generalized Kepler-Coulomb system}
\label{subsec4q}

Let us consider the quantum gKC Hamiltonian:
\begin{equation}
\hat{H} =\frac{1}{2}\biggl(\hbp^2+\sum_{j=1}^n \frac{a_j}{\hat{x}_j^2}\biggl)- \frac{\mu}{|\hbx|} = \frac{1}{2}\biggl(-\hbar^2 \Delta+\sum_{j=1}^n \frac{a_j}{x_j^2}\biggl)-\frac{\mu}{|\bx|} \, .
\label{qgKC}
\end{equation}
Besides the quantum integrals $\hat{Q}_{ij}$ (or  $\hat{P}_{ij}$) given in \eqref{genqRacah}, as for the classical case this system is endowed with the following  additional $n$ fourth-order quantum integrals, the quantum analog of \eqref{gc}, which read:
\begin{align}
\hat{R}_i=\biggl[\sum_{j=1}^{n} \biggl(\frac{1}{2}\{\hat{L}_{ij},  \hat{p}_j\}+\hat{x}_i \frac{a_j}{\hat{x}_j^2}\biggl)-\hat{x}_i \frac{\mu}{|\hbx|} \biggl]^2+\frac{3}{64}\frac{a_i}{\hat{x}_i^2}\biggl(\sum_{j=1}^n \{\hat{x}_j,\hat{p}_j\}\biggl)^2&+\frac{5}{32} \biggl(\sum_{j=1}^n \{\hat{x}_j,\hat{p}_j\}\biggl)\frac{a_i}{\hat{x}_i^2}\biggl(\sum_{j=1}^n \{\hat{x}_j,\hat{p}_j\}\biggl) \nonumber \\
&+\frac{3}{64} \biggl(\sum_{j=1}^n \{\hat{x}_j,\hat{p}_j\}\biggl)^2\frac{a_i}{\hat{x}_i^2}\qquad (i=1, \dots, n) \, .
\label{qfoc}
\end{align}

In the quantum case, the functional relation \eqref{funrel} appears to be:
\begin{equation}
\sum_{i=1}^n \hat{R}_i = -8 \hat{H} \biggl(\hat{C}^{[n]}-\hbar^2 \frac{(2n+1)}{16}\biggl) +\, \mu^2\, .
\end{equation} 
It reduces to the classical relation \eqref{funrel} in the $\hbar \to 0$ limit and to the functional relation characterizing the $nD$ quantum KC system when  all the non-central terms do not appear in the Hamiltonian. In fact, in this case, the total Casimir and the fourth-order quantum integrals read:
\begin{equation}
\hat{C}^{[n]}=-\frac{1}{4} \biggl(\sum_{1 \leq i <j}^n \hat{L}_{ij}^2+\hbar^2 \frac{n(n-4)}{4}\biggl) = -\frac{1}{4} \biggl(\hat{\boldsymbol{L}}^2+\hbar^2 \frac{n(n-4)}{4}\biggl) \, , \qquad\hat{R}_i =\biggl[\frac{1}{2}\sum_{j=1}^{n} \{\hat{L}_{ij},  \hat{p}_j\}-\hat{x}_i \frac{\mu}{|\hbx|} \biggl]^2 = \hat{A}_i^2 \, ,
\label{eq:ctot}
\end{equation}
and the above functional relation collapses to:
\begin{equation}
\sum_{i=1}^n \hat{A}_i^2 = 2 \hat{H}_{\textsf{KC}} \biggl(\hat{\boldsymbol{L}}^2+\hbar^2 \frac{(n-1)^2}{4}\biggl) +\, \mu^2\, , \qquad \biggl(\hat{H}_{\textsf{KC}}=\frac{\hbp^2}{2}- \frac{\mu}{|\hbx|}\biggl) \, .
\label{frelKC}
\end{equation}

Let us now introduce the quantum analog of \eqref{GKCij}, namely the two indices  quantum generators: \begin{equation}
\hat{G}_{ij} :=\frac{1}{2 \imath \hbar}[\hat{R}_i, \hat{P}_{ij}]  \qquad (i \neq j)
\label{GqKCij}
\end{equation}
then, for $i,j,k,l,m \in \{1,\dots, n\}$ all different, we get: 
\begin{equation}
[\hat{R}_i, \hat{H}]=[\hat{P}_{ij}, \hat{H}]=[\hat{G}_{ij},\hat{H}]=0 \, , \qquad  [\hat{R}_i, \hat{P}_{jk}]=0 \, ,
\label{eq:qrelKC}
\end{equation}
and, together with the defining relations of the Racah algebra $R(n)$  given in \eqref{qracah1}-\eqref{qracah5},  we close a quadratic algebra whose additional defining relations are:

\begin{align}
&[\hat{R}_{i}, \hat{P}_{ij}]=:2 \imath \hbar \hat{G}_{ij}\hskip 5.5cm (\hat{G}_{ji}=-\hat{G}_{ij}) \\
&[\hat{R}_i,\hat{R}_j]=-16 \imath \hbar \hat{H} \hat{G}_{ij} \\
&[\hat{R}_i, \hat{G}_{ij}]=-\frac{\imath \hbar}{2}\bigl( \{\hat{R}_i,\hat{R}_j\}-8 \hat{H}\{\hat{R}_i,\hat{P}_{ij}\}+32  \hat{H} \hat{C}_i \hat{R}_j -2 \hbar^2(\hat{H} \hat{R}_i+\hat{H} \hat{R}_j+16  \hat{H}^2 \hat{P}_{ij}+16 \hat{H}^2 \hat{C}_i)+2 \hbar^4 \hat{H}^2\bigl)  \\
&[\hat{R}_i,\hat{G}_{jk}]=8 \imath \hbar  \bigl(\hat{H}  \hat{R}_j \hat{P}_{ik}- \hat{H} \hat{R}_k \hat{P}_{ij} +\hbar^2 (\hat{H}^2 \hat{P}_{ij}- \hat{H}^2 \hat{P}_{ik})\bigl)\\
&[\hat{P}_{ij}, \hat{G}_{ij}]=\frac{\imath \hbar}{2}\bigl(\{\hat{R}_j, \hat{P}_{ij}\}-\{\hat{R}_i, \hat{P}_{ij}\}+4 \hat{C}_i \hat{R}_j-4 \hat{C}_j \hat{R}_i- 4\hbar^2 (\hat{C}_i-\hat{C}_j)\hat{H}\bigl)\\
&[\hat{P}_{ij}, \hat{G}_{ik}]=\imath \hbar \bigl(\hat{R}_j \hat{P}_{ik}-\hat{R}_i \hat{P}_{jk}+\hbar^2(\hat{H}\hat{P}_{jk}-\hat{H}\hat{P}_{ik})\bigl)\\
&[\hat{P}_{ij},\hat{G}_{kl}]=0\\
&[\hat{R}_i, \hat{F}_{ijk}]=\imath \hbar \bigl(\hat{R}_k \hat{P}_{ij}-\hat{R}_j \hat{P}_{ik}+\hbar^2(\hat{H}\hat{P}_{ik}-\hat{H}\hat{P}_{ij})\bigl)\\
&[\hat{R}_i, \hat{F}_{jkl}]=0\\
&[\hat{G}_{ij},\hat{G}_{ik}]=\frac{\imath \hbar}{2} \bigl(\{\hat{G}_{jk}, \hat{R}_{i}\}+8\hat{H}\{\hat{F}_{ijk}, \hat{R}_i\}-2\hbar^2 \hat{H} \hat{G}_{jk}-16 \hbar^2 \hat{H}^2 \hat{F}_{ijk}\bigl) \\
&[\hat{G}_{ij}, \hat{G}_{kl}]=8 \imath \hbar \bigl(\hat{H} \hat{R}_j \hat{F}_{ikl}-\hat{H} \hat{R}_i \hat{F}_{jkl}-\hbar^2 \hat{H}^2 \hat{F}_{ikl}+\hbar^2 \hat{H}^2 \hat{F}_{jkl}\bigl)\\
&[\hat{G}_{ij}, \hat{F}_{ijk}]=- \frac{\imath \hbar}{2}\bigl(\{\hat{P}_{ij}, \hat{G}_{ik}\}+\{\hat{P}_{ij}, \hat{G}_{jk}\}\bigl)\\
&[\hat{G}_{ij}, \hat{F}_{jkl}]= \frac{\imath \hbar}{2}\bigl(\{\hat{P}_{jk}, \hat{G}_{il}\}-\{\hat{P}_{jl}, \hat{G}_{ik}\}\bigl)\\
&[\hat{G}_{ij}, \hat{F}_{klm}]=0 \, .
\end{align}
The above quadratic algebraic structure can be presented in a more compact form if we introduce the new fourth-order operators $\hat{\mathcal{R}}_i:= \hat{R}_i-\hbar^2 \hat{H}$ ($i=1, \dots, n$). In this way, in fact, it reduces to:

\begin{align}
&[\hat{\mathcal{R}}_{i},\hat{P}_{ij}]=:2 \imath \hbar \hat{G}_{ij}\hskip 5.5cm (\hat{G}_{ji}=-\hat{G}_{ij}) \\
&[\hat{\mathcal{R}}_i,\hat{\mathcal{R}}_j]=-16 \imath \hbar \hat{H} \hat{G}_{ij} \\
&[\hat{\mathcal{R}}_i, \hat{G}_{ij}]=-\frac{\imath \hbar}{2}\bigl( \{\hat{\mathcal{R}}_i,\hat{\mathcal{R}}_j\}-8 \hat{H}\{\hat{\mathcal{R}}_i,P_{ij}\}+32  \hat{H} \hat{C}_i \hat{\mathcal{R}}_j -48 \hbar^2 \hat{H}^2 \hat{P}_{ij}\bigl)  \\
&[\hat{\mathcal{R}}_i, \hat{G}_{jk}]=8 \imath \hbar  \bigl(  \hat{H}\hat{\mathcal{R}}_j \hat{P}_{ik}-  \hat{H}\hat{\mathcal{R}}_k \hat{P}_{ij} \bigl)\\
&[\hat{P}_{ij},\hat{G}_{ij}]=\frac{\imath \hbar}{2}\bigl(\{\hat{\mathcal{R}}_j, \hat{P}_{ij}\}-\{\hat{\mathcal{R}}_i, \hat{P}_{ij}\}+4 \hat{C}_i \hat{\mathcal{R}}_j-4 \hat{C}_j \hat{\mathcal{R}}_i\bigl)\\
&[\hat{P}_{ij}, \hat{G}_{ik}]=\imath \hbar \bigl(\hat{\mathcal{R}}_j \hat{P}_{ik}-\hat{\mathcal{R}}_i \hat{P}_{jk}\bigl)\\
&[\hat{P}_{ij},\hat{G}_{kl}]=0\\
&[\hat{\mathcal{R}}_i, \hat{F}_{ijk}]=\imath \hbar \bigl(\hat{\mathcal{R}}_k \hat{P}_{ij}-\hat{\mathcal{R}}_j \hat{P}_{ik}\bigl)\\
&[\hat{\mathcal{R}}_i, \hat{F}_{jkl}]=0\\
&[\hat{G}_{ij}, \hat{G}_{ik}]=\frac{\imath \hbar}{2} \bigl(\{\hat{G}_{jk}, \hat{\mathcal{R}}_{i}\}+8\hat{H}\{\hat{F}_{ijk}, \hat{\mathcal{R}}_i\}\bigl) \\
&[\hat{G}_{ij}, \hat{G}_{kl}]=8 \imath \hbar  \bigl(\hat{H} \hat{\mathcal{R}}_j \hat{F}_{ikl}-\hat{H} \hat{\mathcal{R}}_i \hat{F}_{jkl}\bigl)\\
&[\hat{G}_{ij}, \hat{F}_{ijk}]=- \frac{\imath \hbar}{2}\bigl(\{\hat{P}_{ij}, \hat{G}_{ik}\}+\{\hat{P}_{ij}, \hat{G}_{jk}\}\bigl)\\
&[\hat{G}_{ij}, \hat{F}_{jkl}]= \frac{\imath \hbar}{2}\bigl(\{\hat{P}_{jk}, \hat{G}_{il}\}-\{\hat{P}_{jl}, \hat{G}_{ik}\}\bigl)\\
&[\hat{G}_{ij}, \hat{F}_{klm}]=0 \, .
\end{align}

In complete analogy with the classical case, all the generators cannot be algebraically independent. However, due to the coalgebra symmetry of the model, it is possible to extrapolate the same set of $2n-1$ algebraically independent quantum integrals. Once again, each of the $n$ sets $V_i =\{\hat{C}_{[m]}\} \cup \{\hat{C}^{[m]}\}  \cup \{\hat{H}\} \cup \{\hat{R}_i\}$, at a fixed index $i =1, \dots, n$, turns out to be composed by $2n-1$ algebraically independent quantum integrals, $2n-3$ of them given by the left and right Casimirs of the coalgebra together with the total Casimir $\hat{C}_{[n]} =\hat{C}^{[n]}$.  Again, the closure of the algebra is ensured by the following additional high-order relations among the above quantum generators:
\begin{align}
&\hat{G}_{ij}^2-\hat{C}_i \hat{\mathcal{R}}_j^2-\hat{C}_j \hat{\mathcal{R}}_i^2+\frac{1}{6}\{\hat{P}_{ij}, \hat{\mathcal{R}}_i, \hat{\mathcal{R}}_j\}+\frac{\hbar^2}{3}\bigl(\{\hat{\mathcal{R}}_i,\hat{\mathcal{R}}_j\}+16 \hat{H}\hat{\mathcal{R}}_i(\hat{P}_{ij}+\hat{C}_j)+16 \hat{H} \hat{\mathcal{R}}_j (\hat{P}_{ij}+\hat{C}_i)-36 \hat{H}^2(\hat{P}_{ij}^2-4\hat{C}_i\hat{C}_j)\bigl) \nonumber\\
&\hskip 0.6cm+8 \hbar^4\hat{H}^2\hat{P}_{ij}=0\\
&\{\hat{G}_{ij},\hat{G}_{jk}\}-\frac{1}{6}(\{\hat{P}_{ij},\hat{\mathcal{R}}_j,\hat{\mathcal{R}}_k\}+\{\hat{P}_{jk},\hat{\mathcal{R}}_i,\hat{\mathcal{R}}_j\}-\{\hat{P}_{ik},\hat{\mathcal{R}}_j,\hat{\mathcal{R}}_j\})+\hat{C}_j\{\hat{\mathcal{R}}_i, \hat{\mathcal{R}}_k\}  - \frac{16}{3} \hbar^2 \bigl(\hat{H}\hat{\mathcal{R}}_i \hat{P}_{jk}+\hat{H} \hat{\mathcal{R}}_j \hat{P}_{ik}+\hat{H} \hat{\mathcal{R}}_k P_{ij}\bigl) \nonumber\\
&\hskip 1.65cm +12 \hbar^2 \bigl(\hat{H}^2 \{\hat{P}_{ij},\hat{P}_{jk}\}-4 \hat{H}^2\hat{C}_j\hat{P}_{ik}\bigl)=0\\
&\{\hat{G}_{ij},\hat{G}_{kl}\}-\frac{1}{6}(\{\hat{P}_{ik},\hat{\mathcal{R}}_j,\hat{\mathcal{R}}_l\}+\{\hat{P}_{jl},\hat{\mathcal{R}}_i,\hat{\mathcal{R}}_k\}-\{\hat{P}_{jk},\hat{\mathcal{R}}_i,\hat{\mathcal{R}}_l\}-\{\hat{P}_{il},\hat{\mathcal{R}}_j,\hat{\mathcal{R}}_k\}) \nonumber \\
&\hskip 1.65cm- 12 \hbar^2(\hat{H}^2 \{\hat{P}_{il}, \hat{P}_{jk}\}-\hat{H}^2\{\hat{P}_{ik}, \hat{P}_{jl}\})=0 \\
&\{\hat{G}_{ij},\hat{F}_{ijk}\}-\frac{1}{6}(\{\hat{\mathcal{R}}_k, \hat{P}_{ij},\hat{P}_{ij}\}-\{\hat{\mathcal{R}}_j, \hat{P}_{ij},\hat{P}_{ik}\}-\{\tilde{R}_i,\hat{P}_{ij},\hat{P}_{jk}\})-\hat{C}_i\{\hat{\mathcal{R}}_j,\hat{P}_{jk}\}-\hat{C}_j\{\hat{\mathcal{R}}_i,\hat{P}_{ik}\}+4\hat{C}_i\hat{C}_j\hat{\mathcal{R}}_k \nonumber \\
&\hskip 1.7cm+\frac{\hbar^2}{3}(\{\hat{\mathcal{R}}_i, \hat{P}_{jk}\}+\{\hat{\mathcal{R}}_{j},\hat{P}_{ik}\}+\{\hat{\mathcal{R}}_{k},\hat{P}_{ij}\})=0\\
&\{\hat{G}_{ij}, \hat{F}_{jkl}\}-\frac{1}{6}(\{\hat{\mathcal{R}}_j, \hat{P}_{il}, \hat{P}_{jk}\}+\{\hat{\mathcal{R}}_k, \hat{P}_{ij},\hat{P}_{jl}\}-\{\hat{\mathcal{R}}_j, \hat{P}_{jl}, \hat{P}_{ik}\}-\{\hat{\mathcal{R}}_l, \hat{P}_{ij},\hat{P}_{jk}\})- \hat{C}_j \{\hat{\mathcal{R}}_{l},\hat{P}_{ik}\}+ \hat{C}_j\{\hat{\mathcal{R}}_k,\hat{P}_{il}\}=0\\
&\{\hat{G}_{ij}, \hat{F}_{klm}\}-\frac{1}{6}(\{\hat{\mathcal{R}}_k, \hat{P}_{im}, \hat{P}_{jl}\}+ \{\hat{\mathcal{R}}_{l},\hat{P}_{jm}, \hat{P}_{ik}\}+\{\hat{\mathcal{R}}_m, \hat{P}_{il}, \hat{P}_{jk}\} \nonumber \\
&\hskip 2.2cm- \{\hat{\mathcal{R}}_k, \hat{P}_{il}, \hat{P}_{jm}\}-\{\hat{\mathcal{R}}_l, \hat{P}_{im}, \hat{P}_{jk}\}-\{\hat{\mathcal{R}}_m, \hat{P}_{jl}, \hat{P}_{ik}\}) =0\, .
\end{align}

The above gives higher-order relations, up to order of 10 as differential operators, of the quadratic symmetry algebra. These high-order relations reduce to \eqref{hocKC1}-\eqref{hocKC6} in the classical $\hbar \to 0$ limit.

\section{Concluding Remarks}
\label{conclusions}

In this paper, we have demonstrated that the higher-rank Racah algebra $R(n)$ plays a much broader role as a wide class of $n$D superintegrable systems contain it as the subalgebra of their complete (polynomial) symmetry algebras. This seems to be a common feature of systems whose Hamiltonians share the same coalgebra symmetry, for which the classical/quantum integrals arise as the left and right Casimir invariants through a given symplectic/differential realisation. Here, the Racah subalgebra generators appear as the building blocks of the $2n-3$ classical/quantum integrals above, the latter leading to quasi-maximal superintegrability.
	
	We have presented two cases of $n$-dimensional superintegrable systems whose complete symmetry algebras were not given previously. We have shown that the integrals of motion of the $n$-dimensional Smorodinsky-Winternitz system and generalized Kepler-Coulomb system form higher rank quadratic algebras. This solves one of the open problem on obtaining the symmetry algebras of the $n$D classical and quantum generalized Kepler-Coulomb systems. It is interesting that even with the presence of quartic and quintic integrals, all the commutation relations close to the (higher-rank) quadratic algebra.
	
	Another main result is the derivation of the high-order algebraic closure relations for the quadratic symmetry algebras. Such higher order relations seem to be a generic feature of the polynomial symmetry algebras of superintegrable systems. They are likely to play an important role in the construction of realizations, Casimir invariants and representation theory of the symmetry algebras.

\section*{Acknowledgement}
IM was supported by Australian Research Council Future Fellowship FT180100099. YZZ was supported by Australian Research Council Discovery Project DP190101529 and National Natural Science Foundation of China (Grant No. 11775177).

\addcontentsline{toc}{chapter}{Bibliography}
\bibliographystyle{utphys}
\bibliography{bibliography}

\end{document}